# Implementing a Protocol Native Managed Cryptocurrency

Peter Mell
*National Institute
of Standards and Technology*
Gaithersburg MD, USA
peter.mell@nist.gov

Aurelien Delaitre
*Prometheus Computing*
New Market MD, USA
aurelien.delaitre@nist.gov

Frederic de Vaulx
*Prometheus Computing*
New Market MD, USA
frederic.devaulx@nist.gov

Philippe Dessauw
*Prometheus Computing*
New Market MD, USA
philippe.dessauw@nist.gov

*Abstract*—Previous work presented a theoretical model based on the implicit Bitcoin specification for how an entity might issue a protocol native cryptocurrency that mimics features of fiat currencies. Protocol native means that it is built into the blockchain platform itself and is not simply a token running on another platform. Novel to this work were mechanisms by which the issuing entity could manage the cryptocurrency but where their power was limited and transparency was enforced by the cryptocurrency being implemented using a publicly mined blockchain. In this work we demonstrate the feasibility of this theoretical model by implementing such a managed cryptocurrency architecture through forking the Bitcoin code base. We discovered that the theoretical model contains several vulnerabilities and security issues that needed to be mitigated. It also contains architectural features that presented significant implementation challenges; some aspects of the proposed changes to the Bitcoin specification were not practical or even workable. In this work we describe how we mitigated the security vulnerabilities and overcame the architectural hurdles to build a working prototype.

## I. INTRODUCTION

The United States National Institute of Standards and Technology developed an architecture for a managed cryptocurrency that has many of the features of electronic fiat currencies and includes a governing entity [1]. It is intended to combine the strengths of both fiat currencies and cryptocurrencies. In doing this, it deviates from the goals of most cryptocurrencies by introducing concepts such as central banking, law enforcement, and identity proofed accounts. It also deviates from a government controlled fiat currency world in denying the currency administrator absolute power over financial controls. It enables a currency administrator to enact policy to create a specific cryptocurrency instance from the architecture, usually with immutable configurations that even the administrator cannot violate. This can promote public trust in the currency since the limits to the administrator's power are immutably recorded on the associated blockchain. The architecture uses a public permissionless blockchain approach whereby the administrator's actions are completely transparent. Furthermore, a public set of miners maintaining the blockchain can prevent the administrator from performing unauthorized actions. At the same time, the cryptocurrency is designed to prevent the public miners from taking control from the administrator or from preventing the administrator's transactions from being processed. This architecture thus creates a 'balance of power' between the administrator and the public miners. Additional features include adding role attributes to cryptocurrency accounts that represent fiat currency entities (e.g., commercial banks, central banks, and law enforcement) such that there is created a tree based hierarchy of nodes with roles for all users of the cryptocurrency.

A major limitation to the approach is that it was presented only as a theoretical architecture. It demonstrated what might be possible to create through modest forks to existing cryptocurrencies, specifically using Bitcoin [2] [3] [4] as an example. The empirical work was limited to proposing changes to the implicit Bitcoin specifications in [5] and [6] to add the features necessary for this 'balance of power' managed cryptocurrency approach. No code was developed and no implementation was tested. The ability of [1] to modify the Bitcoin specification to add the needed features indicated that a managed cryptocurrency might be able to be built through a modest fork of an existing cryptocurrency, but it lacked a proof-of-concept prototype built as a protocol native implementation.

In this work, we set out to build such a prototype as an applied research endeavor. We tested whether or not such a managed cryptocurrency system could be built through modest modifications to the code base of an existing cryptocurrency. In this way we explored how to create a protocol native managed cryptocurrency built into the blockchain platform itself and explore the advantages of this approach. This was non-trivial as we did not simply create a token on top of another cryptocurrency. We also wanted to see if this could be done efficiently, with only a modest amount of programming effort (we scoped using half a person year, in part due to resource constraints). We chose to use Bitcoin since [1] described their theoretical model through proposing changes to Bitcoin. We wanted to discover the complexity of modifying Bitcoin to require identity proofing of accounts, establish accounts with roles, enable law enforcement functions, enable central banking functions, and create and visualize a hierarchy tree of accounts that specifies the scope of control of the various management and law enforcement nodes.

An unattributed quote says that 'theory is when you know everything but nothing works.' Yogi Berra said, 'in theory there is no difference between theory and practice. But, in practice, there is.' We found these statements to be true with regard to our implementation of the theoretical work. We discovered that the theoretical model contains several vulnerabilities and security issues that needed to be mitigated. It also contains architectural features that presented significant implementation challenges; some aspects of the proposed changes to the Bitcoin specification were not practical or even workable. We thus had to augment the material in [1] in order to achieve a functional and secure system, especially in areas such as preserving the balance of power, law enforcement powers, management node powers, bootstrapping the system, and the needed movement of accounts within the node hierarchy (e.g., when an account holder changes their account manager). We also encountered difficulties using the Bitcoin code base which necessitated design changes not foreseen in [1]. However, in the end we discovered that it was possible to modestly modify Bitcoin to implement this 'balance of power' managed cryptocurrency approach and to do it with a relatively low amount of programming effort.

In summary, we showed that the theoretical architecture provided by [1] works and can be implemented efficiently. However, we had to change, refine, and augment the original design in order to make it function. This paper describes these changes and the final prototype implementation which we have made publicly available on GitHub (any mention of commercial products is for information only; it does not imply recommendation or endorsement). Note that due to resource constraints, our prototype is not a full implementation. The largest limitation is that the cryptocurrency policy configuration is static, while the full design in [1] permits dynamic policy changes. While not all features were implemented, the core functionality was enabled to provide confidence that the system could be efficiently constructed.

The rest of this paper is organized as follows. Section II presents the theoretical architecture from [1] and discusses relevant Bitcoin architectural features. Section III discusses the vulnerabilities and security issues we discovered in the architecture. Section IV discusses the architectural hurdles that we had to overcome. Section V outlines how we created our prototype system and Section VI presents the related work. Section VII discusses our future plans for the system and Section VIII concludes.

## II. THEORETICAL ARCHITECTURE

The research in [1] provides an architecture that can be instantiated into a cryptocurrency instance through specifying a specific policy configuration. The policy parameters enable or disable feature sets while specifying parameters for cryptocurrency operation. The financially related parameters are just examples of what could be (e.g., limits on money production) and are not intended to be exhaustive given that the identification of financial controls is a related but separate research area. In this architecture, anyone can create an account, but an account cannot do anything unless it is granted one or more roles. The initial block on the blockchain has a 'genesis transaction' that grants roles to the root administrator account and all future role assignments spring from this initial root account. The root account grants roles to other accounts, and those accounts in turn may grant roles to accounts. This sets up a hierarchy of accounts in a tree structure with the root account (or node) being the most authoritative.

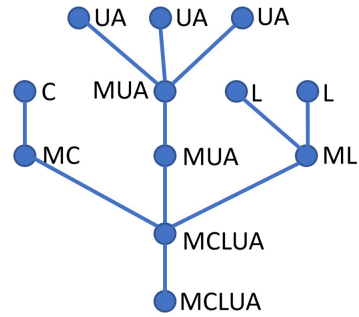

Fig. 1. Example Managed Cryptocurrency Hierarchy (from [1])

The initial root node is given all possible roles so that it can propagate these roles to other accounts. Of particular import is the 'M' currency manager role that enables an account to give its roles to other accounts (or withdraw granted roles) and to modify cryptocurrency policy. Other roles include 'U' user, 'A' account manager, 'C' central banker, and 'L' law enforcement. Their abilities are summarized in [1] as follows:

- 'The U role enables an account to receive and spend coins. An account for which the U role has been removed has its funds frozen.
- The A role enables a node to create accounts with the U role (and only the U role). It may also remove the U label for its descendants.
- The C role enables the creation of new coins (apart from the block mining rewards).
- The L role enables an account to forcibly move funds between accounts, to remove the U label, and to restore a previously removed U label. However, these actions can only be performed against nodes with the same or greater distance from the root.'

Note that in this model the currency administrator controls the root manager node and thus controls the privileges of all other nodes participating in the system. It can thus ensure that the A nodes perform identity proofing of U nodes (if desired). This can enable law enforcement, at least with a court order, to identify individuals within the system. This goes counter to the trend in cryptocurrencies where privacy and non-traceability are key objectives. An example node hierarchy tree with role assignments is shown in Figure 1.

There are three types of transactions that enable accounts with roles to perform their functions: coin transfer mode, role change mode, and policy change mode. A large portion of [1] specifies how to modify the nValue field in Bitcoin (which normally specifies the amount of coin to transfer) to enable the role and policy change functionality while still enabling coin transfer (but now only between accounts with the U role).

Lastly, there are two possible security models. There is an independent mining model where the miners are truly independent from the currency administrator, but they could then as a group deny the inclusion of management transactions

(i.e., role changes and policy changes). This would be similar to a 51 % attack [7] being launched against Bitcoin. To prevent this there is also a dependent mining model where the miners must include a certain number of management transactions every so many blocks. This can prevent a large group of miners from being able to revolt and exclude management transactions as with the independent mining model. However, it shifts the balance of power slightly towards the currency administrator by allowing them to convey a small financial advantage to preferred miners. This risk can be arbitrarily diminished through making certain permanent policy settings.

The theoretical architecture defined in [1] proposed modifying Bitcoin for its implementation. The original Bitcoin whitepaper is available at [2] while detailed explanations can be found in [3], [4], and [5]. Of import to this work is that Bitcoin transfers coins using transactions. The coins are not stored in user accounts but are linked to the transactions themselves. Thus, each transaction has one or more inputs (Vin fields) that bring unspent coins into the transaction and one or more outputs (Vout fields) that declare who can next spend those coin outputs. As shown in Figure 2, a Vin field from some transaction $x$ brings in an unspent Vout field from some transaction $y$. Figure 3 shows the format of a Bitcoin transaction.

## III. DISCOVERED VULNERABILITIES AND SECURITY ISSUES

We discovered vulnerabilities and security issues in the theoretical architecture that needed to be mitigated in order to implement the prototype system. The vulnerabilities enabled violations of the balance of power, replay attacks, and attacks against miners. The security issues included improper scoping of manager and law enforcement powers as well as insecure bootstrapping for establishing cryptocurrency policy.

### A. Preserving the Balance of Power

The research in [1] contains a 'dependent mining model' where the manager can specify that $x$ number of management transactions must be included within each interval of $y$ blocks. One can set $x$ and $y$ through issuing policy transactions. The idea is that this model forces the miners to periodically include management transactions.

However, we have discovered a vulnerability in which the manager can use this feature to take over the blockchain. The manager can initially set $y$ to be high and wait for the community to fully adopt and use the cryptocurrency. Once a significant amount of value has been invested in the cryptocurrency, the manager can issue a policy transaction changing $y$ to be very low. The manager then could, for example, require management transactions to be issued with every block and only send those management transactions to miners whom they favor or control. The miners receiving those transactions would then not propagate them to other miners, preventing the other miners from mining any blocks (since per policy all blocks would have to contain a management transaction). This way, only miners that the manager favored or controlled could publish blocks and the manager could effectively take over the blockchain with effects similar to that of a 51 % attack [8].

Our mitigation is to simply limit how tightly a manager can set $y$. If the specification and developed code reject policy transactions that set $y$ values below some threshold, then the manager is prevented from using this method to take control of the blockchain. The manager could also voluntarily set a minimum threshold for these values using permanent policy transactions issued by the root manager node in order to create public confidence in the cryptocurrency. Even with minimums set, it should be noted that the manager can still implement this attack periodically, favoring their own miners every $y$ blocks if they refuse to issue management transactions in the intervening blocks. This would give a periodic financial advantage to manager favored miners but would be highly visible to the community and would not result in the manager controlling the blockchain. To minimize the impact of this residual attack possibility, $y$ should be required to be high enough to make the financial advantage minuscule. An alternative is to use the independent mining model discussed in II, but this opens up the possibility for the miners to revolt against the manager.

### B. Preventing Replay Attacks

The research in [1] modifies the Bitcoin transactions to support roles because the architecture requires that all transactions include roles. They are brought into the transaction using a modified Vin field; in Bitcoin Vin fields are only used to bring coin into a transaction. Both uses of the Vin field use the same cryptographic protections and one would assume that they would inherit the same security properties. However, this is not the case and it results in a vulnerability in the architecture.

Since roles are spent like coins but never get used up (since you don't lose a role through using it), they can be spent an infinite number of times. This means that transactions that use a role might be able to be replayed. For the typical transactions also transferring coin (e.g., to pay a transaction fee), this is not a problem as the replayed transaction will be rejected because the coin would already have been spent. However, if the transaction does not involve coin it could be replayed. This might happen if the manager owns miners servers and issues management transactions without transaction fees with the intention that their miners will publish them. In this case, there would be no barriers to performing a replay attack. This might result in a situation where law enforcement unlocks an account but can never securely lock it again because the original unlocking transaction can be replayed by anyone.

There are several possible solutions. One approach is to require that all transactions pay some transaction fee while requiring transaction signatures to sign the entire transaction. In our attempt to modify Bitcoin as little as possible, our approach was to change the theoretical model to truly spend roles as if they were coin; once spent they can't be spent again. However, whenever we spend a role by including it in a Vin field we also re-create the same role in one of the Vout fields. The effect is that an account keeps a role when it is spent but the transaction containing the active version of their role can change. Probably the most elegant approach would be to implement the architecture using a cryptocurrency with an accounts based model so that roles are not stored within transactions, but instead within a record associated with each account (discussed more below).

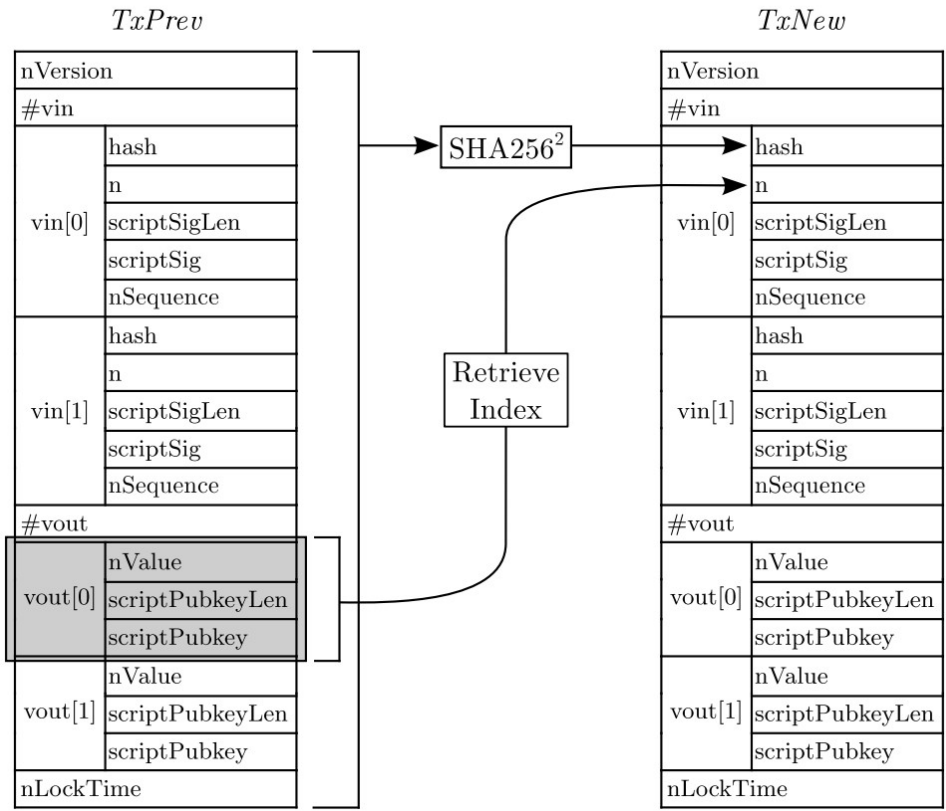

Fig. 2. Bitcoin Vin[] Field Reference to a Previous Transaction (copied from [5]).

| Field name | | Type (Size) | Description |
|---|---|---|---|
| nVersion | | int (4 bytes) | Transaction format version (currently 1). |
| #vin | | VarInt (1-9 bytes) | Number of transaction input entries in *vin*. |
| vin[] | hash | uint256 (32 bytes) | Double-SHA256 hash of a past transaction. |
| | n | uint (4 bytes) | Index of a transaction output within the transaction specified by *hash*. |
| | scriptSigLen | VarInt (1-9 bytes) | Length of *scriptSig* field in bytes. |
| | scriptSig | CScript (Variable) | Script to satisfy spending condition of the transaction output (*hash,n*). |
| | nSequence | uint (4 bytes) | Transaction input sequence number. |
| #vout | | VarInt (1-9 bytes) | Number of transaction output entries in *vout*. |
| vout[] | nValue | int64_t (8 bytes) | Amount of $10^{-8}$ BTC. |
| | scriptPubkeyLen | VarInt (1-9 bytes) | Length of *scriptPubkey* field in bytes. |
| | scriptPubkey | CScript (Variable) | Script specifying conditions under which the transaction output can be claimed. |
| nLockTime | | unsigned int (4 bytes) | Timestamp past which transactions can be replaced before inclusion in block. |

Fig. 3. Bitcoin Transaction Format for Sending Bitcoin (BTC), copied from [5].

## C. Preventing Managers from Attacking Miners

In [1] all accounts must have the U role for them to receive or spend coin. The purpose is to force all participants in the cryptocurrency to be identity proofed by an account manager in order to receive the U role. This in turn supports 'know your customer' laws, which have been a challenge for most cryptocurrencies [9]. However, this also creates a vulnerability. The manager could keep track of the accounts receiving block rewards and remove the U role from those accounts (thus freezing the funds). The public miners would then have no financial incentive to mine and then the manager's own mining servers could take over the majority of mining. This would give the manager the ability to launch a 51 % attack [8] and to a large degree control the blockchain.

Our solution is to enable miners to deposit block rewards into any account, regardless of whether or not it has been registered in the system or has any roles. Also, we handle the coin from these coinbase transactions (the mining reward transactions) specially such that it can be sent without the owning account needing the U role. This prevents the currency administrator from freezing the mining reward coinbase funds. However, once coinbase coin is sent away from the original account it becomes normal coin subject to the normal requirements (it can't be spent without the associated account having the U role).

## D. Scoping Law Enforcement Powers

In [1] law enforcement powers are both too limited and too relaxed. They are too limited in that law enforcement can only lock accounts through removing the U label. Law enforcement nodes can't prevent an account using its other roles (M, C, A, or L). This is a major issue in the event that an account is stolen. On the other hand, law enforcement powers are too relaxed in that law enforcement nodes can effect any node higher in the account hierarchy tree regardless of whether or not it is on the same branch. This effectively gives law enforcement nodes a global reach (which is especially problematic if a law enforcement node is compromised).

Our solution is to reflect account locking not through the removal of the U role but by setting a locked flag. We use one of the unused bits in the nValue field for role change mode to set this flag. If the flag is set it temporarily disables all roles, not just the U role. This stops all activity by the targeted account, giving law enforcement the powers it needs to freeze stolen accounts. At the same time, we put additional restrictions on law enforcement nodes by only giving them authority over nodes farther from the root on the same branch of the node hierarchy tree. More precisely, we define the scope of control of a law enforcement node by traversing backwards until the first node is found with the manager role and then by performing a breadth first search to reveal all nodes within scope. This enables law enforcement nodes to 'hang' off of manager nodes in the tree (they don't have to be inline on each branch).

## E. Management Node Powers

In [1] management nodes also had powers that were too relaxed. They were required to have any role that they would want to grant. This resulted in management nodes having powers that they had no intention of using. Also, their scope of control was the same as law enforcement giving each M node low down in the hierarchy tree an almost global reach.

Our solution was to limit their scope to nodes reachable by breadth first search and to limit management nodes to only having the M role. However, in our approach management nodes can add any role to other nodes. This gives more power to a manager node (which might be seen as decreasing security) but it limits that power to a more narrow scope creating what we believe is a rational compromise.

## F. Policy Bootstrapping

In [1], it is not stated how the initial policy is defined for an instantiated cryptocurrency. It is implied that some configuration file, apart from the blockchain, must exist that provides the original parameter settings. These settings may or may not then be subsequently overridden through policy transactions on the blockchain. The result may be that some policy is defined on the blockchain and some through an original configuration file. Given that the configuration file wouldn't have the same cryptographic protections as blockchain transactions, the distributor of the node software for maintaining the blockchain could conceivably change policy using software updates through modifying the configuration file.

Our solution is to eliminate the need for the unsecured initial configuration file. We do this by specifying that all policy is initially defined as permissive as possible. We then require that all policy parameters be defined explicitly on the blockchain within the first $x$ blocks (as defined in the full node software distribution). Thus early in the blockchain, ideally prior to it being released publicly, the manager will have to explicitly record all possible policy parameters within cryptographically secured blocks.

We also discovered that the original root management node should not be used to set the initial policy (except for policy settings intended to be permanent). This is because, per [1], management nodes closer to the root are more authoritative; any root manager node policy decisions will prevent any other management node from changing that policy. Also, the root management node account ideally should never be used after the initial few blocks and its keys should be physically stored in a vault to eliminate the possibility of it being compromised. Thus, if the root node is used to set policy it should only be used to set permanent policy that, by design, will never be changed.

## IV. ARCHITECTURAL CHALLENGES

Apart from mitigating vulnerabilities in the original design, we encountered several architectural challenges where it was not practical or even possible to directly implement the theoretical architecture. In this section we describe the primary challenges, how we modified the theoretical design to overcome them, and how we implemented those changes.

### A. Dual Signature Requirements for Coin Transfer Transactions

In [1] an account must have the U role to both spend and receive coin. It specifies that these roles must be brought into each coin transfer transaction using two separate Vin fields. However, this requires both the sender and receiver to sign the transaction which would require off blockchain coordination and some unspecified infrastructure to support this.

This could be resolved by including the coin transfer recipient only in the Vout field (not the Vin) and requiring full nodes to check the U role on the account listed in the Vout field (without explicitly bringing it into the transaction using a Vin field), at the cost of additional tracking overhead. Our mitigation was to only require the U role for spending coin. Any account then can receive coin, but may not be able to spend it. This results in only a single account needing to sign coin transfer transactions and eliminates additional overhead.

### B. Node Movement

In [1] there is no mention of how accounts can change position within the node hierarchy graph once they have been created. This is necessary, for example, for users that want to use different account managers. Besides moving nodes, edges in the graph may need to be moved in order to cut out compromised nodes but leave the rest of the node hierarchy intact.

To implement the needed functionality, we created the idea that if a node adds roles to an account that has no roles, this creates an edge in the node hierarchy graph from the node adding the roles to the node representing the account gaining the roles. If an edge already existed to the node gaining the roles (which would happen if an account received roles and then deleted them), the prior edge will be deleted in order to preserve the required tree structure.

To prepare a node to be moved, the relevant account can unilaterally remove its own roles or else a manager whose scope covers the node can remove the roles. Using this paradigm, nodes can be moved around the node hierarchy tree. It also doesn't require explicitly coding edge creation and deletion within the modified Bitcoin protocol, which would have been unnecessarily complicated. A drawback is that node movement requires a two step process: one transaction to remove roles and another to add them back in (thus removing the old edge and creating the new edge). In our future work we will design a format where a single transaction does this atomically. Complicating this may be the need for dually signed transactions to prevent security violations (which we are trying to avoid, see section IV-A). Our current two step approach ensures that the role removal, node movement, and edge addition only happens through transactions issued by nodes authorized to perform those activities.

### C. Determining Transaction Types

The theoretical architecture in [1] uses the most significant bits of an nValue field to determine the type of transaction being processed: role change, policy change, or coin transfer. The nValue fields, in the original Bitcoin, specify the amount of coin to be spent. Using the leftmost bits as control bits is conceivably risky because a bug in the code might interpret the leftmost control bits as value bits for moving or create large amounts of coin. More problematic though is that the Bitcoin implementation uses the leftmost bit of the nValue field as a signed bit.

For these reasons, we chose to deviate from [1] and not use the leftmost bits of the nValue field to determine the type of transaction. Instead, we determined the type of transaction using the transaction version number; this then determines how the nValue fields within a transaction are handled. We created three transaction version numbers, each of which correspond to the three different modes for evaluating nValue fields (role change, policy change, and coin transfer). Lastly, we also changed to using the nValue low order bits for specifying roles and policy change types in case those nValue fields ever got interpreted as coin transfer fields through some bug or attack. This would then limit the damage done by having fewer coin inadvertently transferred or created.

### D. Transaction Fees

Since we determine transaction type (role change, policy change, or coin transfer) through the transaction version number, it means that the mode of all the nValue fields in the Vout fields are determined by that number. However, it is usually necessary to pay a transaction fee for most transactions and there is usually change that must be sent back to the sender. This is not possible then for the role and policy change transactions because the nValue fields of the Vout fields change roles/policies; they don't send coin as in the original Bitcoin specification. We solved this simply by specifying which Vout field is always the change sent back to the originator of the transaction (which may be 0 coin on occasion).

## V. DEVELOPED PROTOTYPE

Our prototype was developed publicly through Github and is available within the project 'usnistgov/managed-cryptocurrencies-bitcoin'. We built our prototype through forking and modifying the C++ Bitcoin codebase available on Github at 'bitcoin/bitcoin'.

For flexibility, efficiency, and portability we ran our modified bitcoin peer-to-peer network for development and testing on a local virtualized environment. For our testing, we thus had a single virtual machine (VM) executing the entire distributed Bitcoin network. We used the Vagrant virtual machine manager with Virtualbox as the VM provider. Within the VM, we used the Docker Engine to run a set of containers to represent the nodes on the modified Bitcoin network. This enabled us to simultaneously run five Bitcoin miners within a single VM to maintain our test blockchain. Note that we artificially reduced the mining difficulty to enable quick block production for testing and demonstration purposes. Lastly, we used the GraphViz library to enable us to visualize the node hierarchy tree. To make access control decisions for role and policy change transactions, it was inefficient to look up individual node roles using the tree. Thus, we separately maintained an associative array mapping node names to a list of their roles. The tree was only necessary for determining the scope of control of one node over others (e.g., for the law enforcement and manager nodes).

An example output tree is shown in Figure 4. Within each node in parenthesis is listed the roles activated for that node and its state (locked or unlocked). The labels are deciphered as follows: M-manager, C-central banker, L-law enforcement, U-registered user, A-account manager, D-disabled account) Node 0 is the root node created in the genesis block. It should normally never be used directly for security reasons and so Node 1 was created as the 'active' manager. Node 3 is the central banker; it could have hung off of Node 1 but it was useful for our example to have it as a child under Node 0. Node 2 is law enforcement with the scope of all that is reachable from Node 1 (all nodes except 0, 3, and 11). Nodes 4 and 5 are account managers. Node 6 is a user account that has been disabled by law enforcement. Nodes 7, 8, and 10 are ordinary users. Node 9 is a node who has had all its roles removed (either done by Node 9 itself, its account manager Node 5, or one of the manager nodes 0 or 1). This might have been done because Node 9 was compromised or because it is being prepared to move to another part of the tree under a different account manager. Node 11 is a node that has been active in the cryptocurrency but has no roles and has never had any roles (due to their being no edge to it). It represents an account created by a miner to store coinbase coins, that can be spent without needing any roles.

## VI. RELATED WORK

To our knowledge, [1] is the only work proposing a managed cryptocurrency that has a balance of power where the public can hold the manager accountable. There have been many government cryptocurrencies proposed but these differ in that they are often not managed, don't use roles, or don't have a balance of power.

Multichain [10] is a system that might appear to be similar in that it contains management features. However, Multichain enables a permissioned chain where what is managed is which entities have the privilege of mining. This is opposite of our prototype that enables open mining. That said, we may explore modifying Multichain to implement [1] while leveraging a permissioned chain whose membership is defined by the current members (not the manager).

There are many government cryptocurrencies proposed and in development (for example [11], more citations are in [1]). However, none of these have yet come to fruition except the Venezuelan Petro [12], which to our knowledge is the only existing government issued cryptocurrency.

There is research proposing a Fedcoin [13], a cryptocurrency that would support central banks with a permissioned blockchain that complies with 'know your customer' laws [9]. It is based on RS—Coin [14], one of many cryptocurrencies advertised to support central and commercial banks with international transaction handling. Others argue that central banks don't need a cryptocurrency, but instead a new form of electronic money [15]. There are also concerns with the amount of power a government could leverage through creating a Fedcoin [16].

## VII. FUTURE WORK

There are two major changes to be made in future iterations of the implementation: using an account model and better handling of compromised nodes.

### A. Using an Account Model

Bitcoin uses an unspent transaction output (UTXO) model. Coin is not stored within user accounts but within the transactions themselves. All transactions have outputs (representing coin) and any unspent output may be spent by another transaction. Who may spend a given output is determined by a script that usually specifies the public key of a particular account. There is no single data structure on the blockchain that shows the coin associated with a particular account.

This works well for Bitcoin, but immediately became awkward for the implementation of our managed cryptocurrency prototype. In the theoretical architecture, accounts have roles that specify their privileges in the system and these roles are specified in nValue fields. Without a central data structure for each account, the roles had to be treated like coin and be spent repeatedly as an account used those roles. In our system, an account's roles are transaction outputs and the active copy (the one that hasn't yet been spent) is temporarily in one particular transaction. We simplified this, compared to the theoretical architecture, by requiring that any role additions and removals repeat the remaining roles. Thus, all of an account's roles are always designated within a single transaction, not spread out among many transactions as would have occurred through a direct implementation of the theoretical architecture.

Our future approach will be to implement the system through forking cryptocurrency code that uses an account model instead of an UTXO model. This is possible because the theoretical architecture is not tied to any particular cryptocurrency. A likely candidate replacement cryptocurrency would be Ethereum due to its maturity, but this choice would bring in the added complexity of a codebase that supports smart contracts. A mature Bitcoin-like cryptocurrency without smart contract capabilities that uses an account model might be better suited.

### B. Handling Compromised Nodes

In section III-D we expand the law enforcement powers to disable all the roles of an account to handle the case where a node is compromised (in [1] only the ability to send and receive coin was disabled). However, this does not allow the compromised node to be recovered. To do this, we propose that all nodes should have two sets of cryptographic key pairs. The first set is used for the daily signing of transactions for the associated account. The second set is stored offline and is used only to replace the first set. This enables account owners to unilaterally re-establish control over their accounts without having to involve a manager node (one with the M or A role). However, it will require the development and implementation of a new transaction type to enable this resetting of the first key pair.

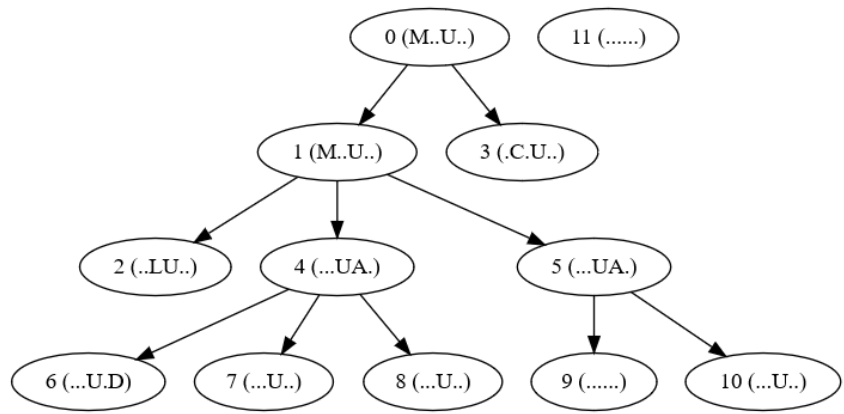

Fig. 4. Example Output Showing a Node Hierarchy.

## VIII. Conclusion

The theoretical managed cryptocurrency architecture proposed in [1] can be efficiently developed from an existing cryptocurrency codebase and deployed (despite the many implementation issues that had to be overcome). An important result of this is that we have shown that the novel balance of power concept, whereby a manager and public miners jointly control a cryptocurrency, is a feasible mechanism to be explored for future cryptocurrencies. Another result of our work is to show the practicability of adding roles to cryptocurrency accounts and the capabilities that can be achieved through these roles (in particular for mimicking fiat currency mechanisms). Lastly, we note that building such a protocol native managed cryptocurrency within a blockchain platform itself was non-trivial but we showed that it could be accomplished with only a modest cost in programming effort.

In summary, we have shown that the theoretical system in [1] can be implemented in such a way as to not just leverage many of the strengths of modern cryptocurrencies, but also leverage the capabilities of traditional fiat currencies. While this goes against the goals and directions of most cryptocurrency efforts which are promoting greater privacy and autonomy from managing institutions, this result may be useful for large institutions (e.g., governments) investigating future electronic currency approaches. We do not necessarily believe that the architecture in [1] provides the answer for such a use case, but it and our applied research in this work may open up new research directions to better support large institutions issuing their own managed cryptocurrencies.

## References

[1] P. Mell, "Managed blockchain based cryptocurrencies with consensus enforced rules and transparency," in *2018 17th IEEE International Conference On Trust, Security And Privacy In Computing And Communications/12th IEEE International Conference On Big Data Science And Engineering (TrustCom/BigDataSE)*.  IEEE, 2018, pp. 1287–1296.
[2] S. Nakamoto, "Bitcoin: A peer-to-peer electronic cash system," 2008. [Online]. Available: https://bitcoin.org/bitcoin.pdf
[3] J. Bonneau, A. Miller, J. Clark, A. Narayanan, J. A. Kroll, and E. W. Felten, "Sok: Research perspectives and challenges for bitcoin and cryptocurrencies," in *Security and Privacy (SP), 2015 IEEE Symposium on*.  IEEE, 2015, pp. 104–121.
[4] A. Narayanan, J. Bonneau, E. Felten, A. Miller, and S. Goldfeder, *Bitcoin and Cryptocurrency Technologies: A Comprehensive Introduction*. Princeton University Press, 2016.
[5] K. Okupski, "Bitcoin developer reference," 2016. [Online]. Available: https://lopp.net/pdf/Bitcoin_Developer_Reference.pdf
[6] "bitcoinwiki protocol documentation," accessed: 2017-12-29. [Online]. Available: https://en.bitcoin.it/wiki/Protocol_documentation
[7] J. Yli-Huumo, D. Ko, S. Choi, S. Park, and K. Smolander, "Where is current research on blockchain technology? a systematic review," *PloS one*, vol. 11, no. 10, p. e0163477, 2016.
[8] S. Barber, X. Boyen, E. Shi, and E. Uzun, "Bitter to betterhow to make bitcoin a better currency," in *International Conference on Financial Cryptography and Data Security*.  Springer, 2012, pp. 399–414.
[9] M. Staples, S. Chen, S. Falamaki, A. Ponomarev, P. Rimba, A. Tran, I. Weber, X. Xu, and J. Zhu, "Risks and opportunities for systems using blockchain and smart contracts," 2017. [Online]. Available: https://publications.csiro.au/rpr/download?pid=csiro:EP175103dsid=DS2
[10] G. Greenspan, "Multichain private blockchainwhite paper," 2015. [Online]. Available: https://www.multichain.com/download/MultiChain-White-Paper.pdf
[11] L. Coleman. An inside look at chinas government controlled cryptocurrency project. [Online]. Available: https://www.ccn.com/an-inside-look-at-chinas-government-controlled-cryptocurrency-project
[12] D. B. Alexandra Ulmer, "Enter the 'petro': Venezuela to launch oil-backed cryptocurrency," Reuters, Dec. 2017.
[13] S. Gupta, P. Lauppe, and S. Ravishankar, "A blockchain-backed central bank cryptocurrency," 2017. [Online]. Available: https://zoo.cs.yale.edu/classes/cs490/16-17b/gupta.sahil.sg687
[14] G. Danezis and S. Meiklejohn, "Centrally banked cryptocurrencies," *arXiv preprint arXiv:1505.06895*, 2015.
[15] A. Berentsen and F. Schar, "The case for central bank electronic money and the non-case for central bank cryptocurrencies," 2018. [Online]. Available: https://doi.org/10.20955/r.2018.97-106
[16] T. Aube. The terrifying future of fedcoin. [Online]. Available: https://hackernoon.com/the-terrifying-future-of-fedcoin-ddcbef2b9592